\documentclass[sigconf]{acmart}

\usepackage{amsmath}  
\usepackage[mathscr]{eucal}
\usepackage{graphicx}
\usepackage{enumitem}
\AtBeginDocument{%
  \providecommand\BibTeX{{%
    \normalfont B\kern-0.5em{\scshape i\kern-0.25em b}\kern-0.8em\TeX}}}

\setcopyright{rightsretained}
\acmConference[SIGIR eCom'23]{ACM SIGIR Workshop on eCommerce}{July 27, 2023}{ Taipei, Taiwan}
\acmYear{2023}
\copyrightyear{2023}
\makeatletter
\renewcommand\@formatdoi[1]{\ignorespaces}
\makeatother
\acmISBN{}

\acmPrice{15.00}
\acmISBN{978-1-4503-XXXX-X/18/06}

\begin{document}

\title[Product Search as Program Synthesis]{(Vector) Space is \textit{Not} the Final Frontier:\\Product Search as Program Synthesis}

\author{Jacopo Tagliabue}
\email{jacopo.tagliabue@nyu.edu}
\affiliation{%
  \institution{New York University, Bauplan}
  \city{New York City}
  \state{NY}
  \country{USA}
}

\author{Ciro Greco}
\email{ciro.greco@bauplanlabs.com}
\affiliation{%
  \institution{Bauplan}
  \city{New York City}
  \state{NY}
  \country{USA}
}


\renewcommand{\shortauthors}{Tagliabue et al.}

\begin{abstract}
As ecommerce continues growing, huge investments in ML and NLP for Information Retrieval are following. While the vector space model dominated retrieval modelling in product search -- even as vectorization itself greatly changed with the advent of deep learning --, our position paper argues in a contrarian fashion that program synthesis provides significant advantages for many queries and a significant number of players in the market. We detail the industry significance of the proposed approach, sketch implementation details, and address common objections drawing from our experience building a similar system at \textit{Tooso}.
\end{abstract}



\keywords{product search, semantic parsing, program synthesis, large language models}

\begin{teaserfigure}
  \includegraphics[width=\textwidth]{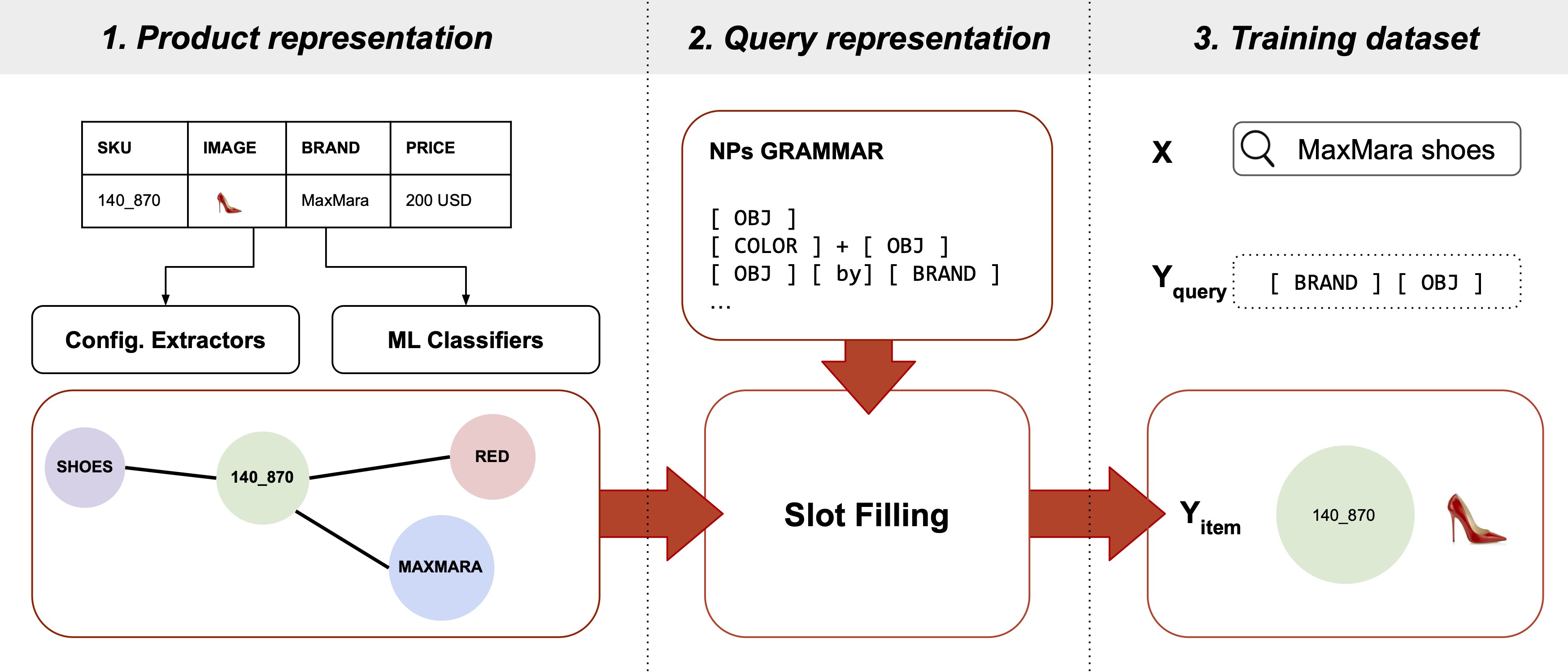}
\caption{How to build a query parser with no manual annotations (Section \ref{sec:parsing}). (1): extractor for product features from the catalog; (2): a grammar covering common logical forms; (3) dataset generated with queries and matching semantic form, to train a parser for runtime query analysis.}
\label{fig:teaser}
\end{teaserfigure}


\maketitle

\section{Introduction}
\label{sec:intro}

\begin{quote}
``Now, like all great plans, my strategy is so simple an idiot could have devised it'' --~\textit{Zapp Brannigan}
\end{quote}

The	explosive growth of ecommerce~\cite{emarketer2020} brought equally impressive innovation in Information Retrieval (IR)~\cite{10.1145/3459637.3482276}, with \textit{product search} now to representing 30\% to 60\% of total online revenues~\cite{bigCommerce2020,retailDive2020,forrester2021}. Building on decades of literature in web and document retrieval, product search is typically modelled as a two-step process: \textit{candidate selection} (\textit{retrieval} \cite{10.1145/361219.361220}) and \textit{re-ranking}~\cite{mitra2018an,choudhary2022anthem,Pei2019PersonalizedCR,Li2019FromSR}. The most widespread model for retrieval is the \textit{vector space model} (VSM) \cite{gillick2018end,Karpukhin2020DensePR,tay2022transformer}, according to which relevance is approximated by the distance between a query vector and a product vector in a suitable space. Even as deep learning drastically altered vectorization \cite{Nogueira2019MultiStageDR}, it did not call into question the tenets of the VSM, or the idea that re-ranking is needed to push down the page irrelevant items wrongfully retrieved \cite{Yan2018APD,10.1145/2911451.2926725}. It is important to remember that most real-world search engines leverage VSM in one form or another: sparse BM25 retrieval in Elasticsearch may be implemented very differently from dense retrieval on Redis Vector Search\footnote{\url{https://redis.io/docs/stack/search/reference/vectors/}}, but they all share the core idea of VSM. Namely, that retrieval is fundamentally approximated by distance in a vector space.

We argue that \textit{program synthesis through semantic parsing} provides a principled and viable alternative to VSM for product search. In this perspective, search queries are (informal) instructions for knowledge bases, as opposed to points in a vector space \footnote{As we explain below (Fig.~\ref{fig:search}, our approach is to parse a search query to an intermediate semantic representation, and then translates the latter into a program, handling the shopping query ``as if it were instructions''; program synthesis may also be construed directly from natural language \cite{inproceedings}. We will refer to parsing and synthesis somehow liberally below, since it's clear how to move from one to the other.}. We shall defend two main claims: 

\begin{enumerate}
    \item VSM is an \textit{indirect} representation of \textit{meaning} that is necessary for large unstructured documents, such as those in web search; however, under different circumstances, where search queries are interpreted against product catalogs, \textit{direct} representation is feasible and useful;
    \item explicit representations unlock a powerful search experience where formal inferences can be made to improve retrieval, while ranking is used as a device for personalization.
\end{enumerate}

Historically, ecommerce tech has been focusing mostly on the challenges of big players, while a larger market share represented by mid-to-large websites has been neglected \cite{10.1145/3460231.3474604}. While we recognize the intrinsic limits of \textit{position} papers, we believe our contrarian argument will benefit from the freedom allowed by this format. Our arguments proceed as follows: we first establish some empirical facts about ecommerce search at the ``Reasonable Scale''; we then showcase the virtues of program synthesis, \textit{assuming} a semantic oracle. Finally, we show how such a system can actually be built.

We believe this work to be valuable for a broad set of practitioners, solving specific use cases in this segment of the market or working on SaaS solutions\footnote{As a business context about this blooming industry, Algolia and Bloomreach raised >USD200M each in venture money in the last few years ~\cite{AlgoliaRound,BloomreachRound}, and Coveo raised >CAD200M at IPO \cite{CoveoRound}.}. Even if most of the arguments we present are theoretical, these ideas have been successfully implemented in a company before (\textit{Tooso}), and played an important role in its acquisition by a public market leader (\textit{TSX:CVO})\footnote{While most of these ideas have been developed in 2017-2019, we have updated our arguments to reflect the most recent advancements in the field.}. 

\section{An Industry Perspective}
\label{sec:industry}

\begin{quote}
``Hooray! A happy ending for the rich people.'' --~\textit{Dr. Zoidberg}
\end{quote}

\begin{figure}
  \centering
  \includegraphics[width=6.5cm]{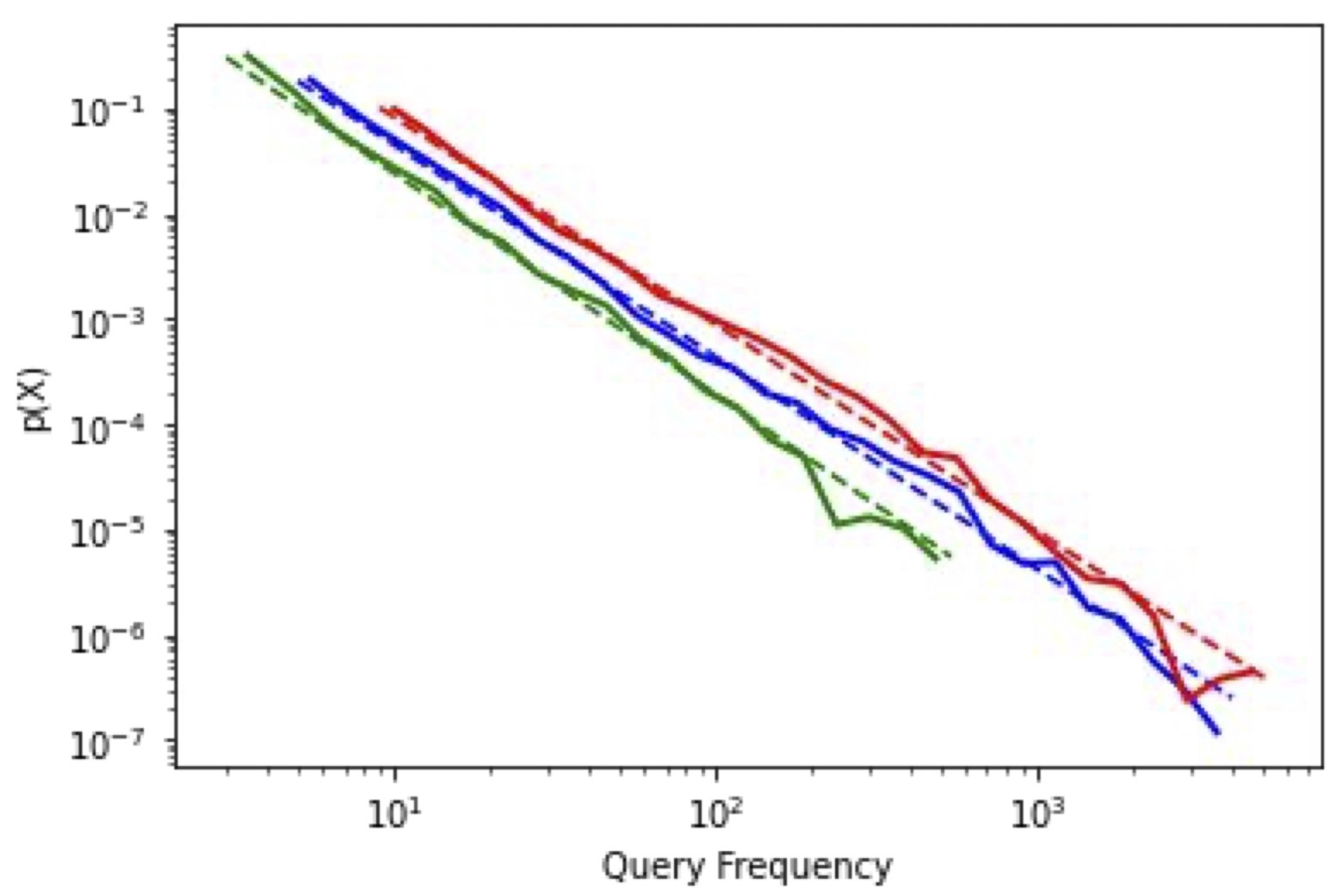}
  \caption{Anonymized query frequency distribution (and fitted power-law PDF) on a log-log plot for three RSc shops in the literature \cite{Requena2020,BianchiSIGIReCom2020,CoveoSIGIR2021}.}
  \label{fig:powerlaws}
  \vspace{-4mm}
\end{figure}

While the idiosyncrasies of product search have been partially documented before \cite{Tsagkias2020ChallengesAR,Brenner2018EndtoEndNR}, most ecommerce systems are still designed from the same building blocks as document search: VSM for retrieval, Machine Learning (ML) for re-ranking using all types of signals. In our experience, the farther you go from planetary scale retailers, the less product search will resemble web search.

Because digital transformation is consistently taking place in the retail industry, most ecommerce search systems are now deployed outside of Big Tech Retailers. We are going to describe the mid-long tail of ecommerce implementations as the ``Reasonable Scale'' (RSc) \cite{towardDataScienceJT,ericMLOPSMess,tagliabue2023reasonable,10.1145/3475965.3479315}.
While RSc is intended to be a loose concept~\cite{10.1145/3460231.3474604}, practitioners typically know it when they see it \cite{Metaflow}. 

\begin{figure}
  \centering
  \includegraphics[width=7.5cm]{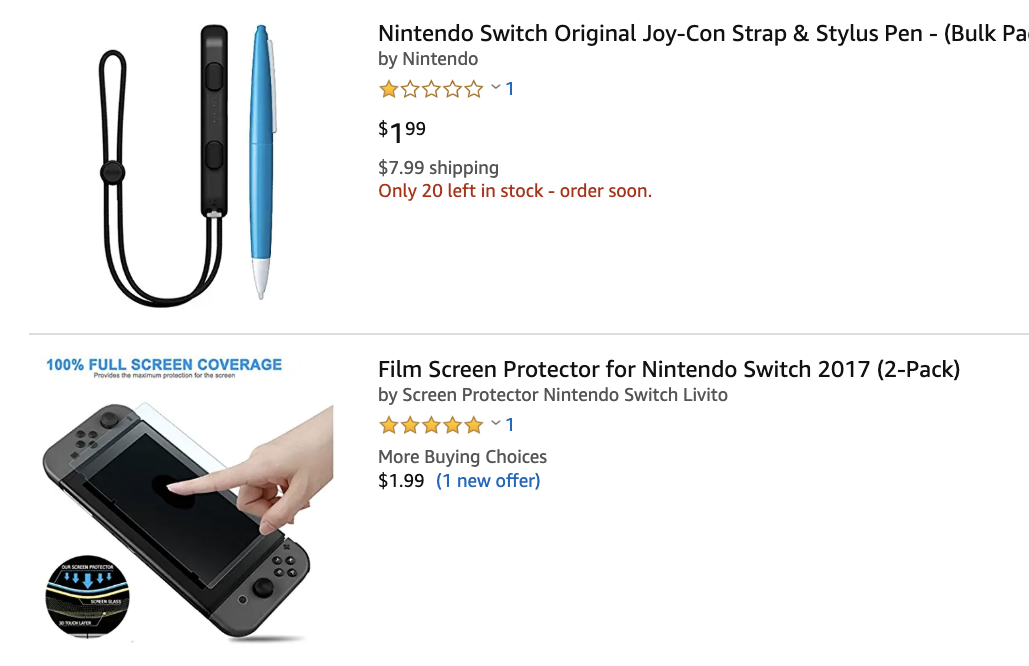}
  \caption{Price re-ordering on~\textit{Amazon.com}, showing degrading relevance in the result set when querying for ``nintendo switch'', and then re-ranking based on price.}
  \label{fig:amazon}
  \vspace{-4mm}
\end{figure}

A number of strategies need to be different at RSc. For instance, instead of several millions of SKUs, RSc shops may have 10K to 100K products and still make >100M USD in yearly revenues. Queries on inventories of this size can easily have result sets of ~10 golden items. In this context, no re-ranking strategy will be able to hide irrelevant products from the user: for the typical strategy of hiding results in page two\footnote{``The best place to hide a dead body is page two of Google.''} to work, there should be a page \textit{two} to begin with. Even as inventory grows, VSM may go against shoppers' preference: for price-sensitive items, users often sort results by price \cite{tagliabue-etal-2020-grow}. When this happens, sub-optimal candidate selection can hurt the experience (Fig.~\ref{fig:amazon})(see also the cases discussed in \cite{King2023} with regard to prices and sizes).

To paint a more quantitative picture of the RSc, we can leverage our unique and privileged position as SaaS practitioners with access to dozens of different real-world deployments. In particular, there are two main facts that turn out to be crucial for our approach (Section \ref{sec:parsing}):

\begin{enumerate}
    \item product search mostly deals with short queries in the form of Noun Phrases (NPs) describing entities and properties (e.g. ``red shoes'' or ``Dell laptop'') \cite{userSearch}. Query examples from RSc shops can also be found in \cite{10.1145/3366424.3386198} (Table 1) and \cite{bianchi-etal-2021-language}; 
    \item a small number of queries account for a significant portion of the distribution, making superior relevance for top queries extremely impactful for the overall experience. In the frequency distribution of a month of anonymous query data sampled from three RSc shops in two languages, the top 1-to-5\% queries account for \textit{half} of the total individual queries (Fig. \ref{fig:powerlaws}). 
\end{enumerate}

The first observation is important as parsing gets harder with longer queries; the second observation is important as it indicates how to align technological objective with business outcomes -- i.e., solving parsing for short queries is a very good place to start.

Taken together, they both re-affirm the peculiarities of product search, but from a novel and unusual angle: interestingly, both facts are \textit{not} true for web or big-scale ecommerce search -- as the numbers of users / items get larger and revenues grow into billions, the tail of the query distribution gets both longer and more important. In other words, while the general linguistic behavior for users of \textit{Amazon} or \textit{Facebook} is also be NP-based, the tail is disproportionally more important: the tail is longer, as big catalogs invite a larger set of inputs, and the tail is more valuable, as marginal improvements in rare queries translate in sizable monetary gains. While we believe our approach can be used, under the appropriate circumstances, at any scale, its novelty and impact are more easily noticeable for RSc deployments.

\section{Searching with an oracle}
\label{sec:oracle}

Originally developed for large documents and long queries, VSM is a useful approximation as it provides a retrieval strategy that avoid \textit{explicitly modelling for meaning}, which has long been thought to be an intractable problem: what would be the logical form \cite{Jia2016DataRF} of this Wikipedia page\footnote{\url{https://en.wikipedia.org/wiki/Transformer_(machine_learning_model)}}? As we argue below, the challenges of explicit representations are eased for product search: on the query side, real-world data shows that NP-like queries are very impactful (Section \ref{sec:industry}); on the item side, products are remarkably different from long documents: products are well-defined entities, which can be described through a sortal (i.e. the type of object, e.g. ``shoes'') and few key properties (e.g. color, material, size, brand, price - crucially those more often used by shoppers \cite{bianchi-etal-2021-query2prod2vec, bianchi-etal-2021-language}). In other words, products already come into an IR system as (quasi) \textit{structured} information.  

What would a search-as-parsing experience look like? We first sketch the general experience we have in mind through a ``parsing oracle'' (PO) - i.e. an idealized system that is able to: 

\begin{itemize}
    \item at runtime, return the logical form of a query;
    \item at indexing time, given a product (as contained in a digital catalog \cite{CoveoSIGIR2021}), return its properties.
\end{itemize}

\begin{figure}
  \centering
  \includegraphics[width=7.5cm]{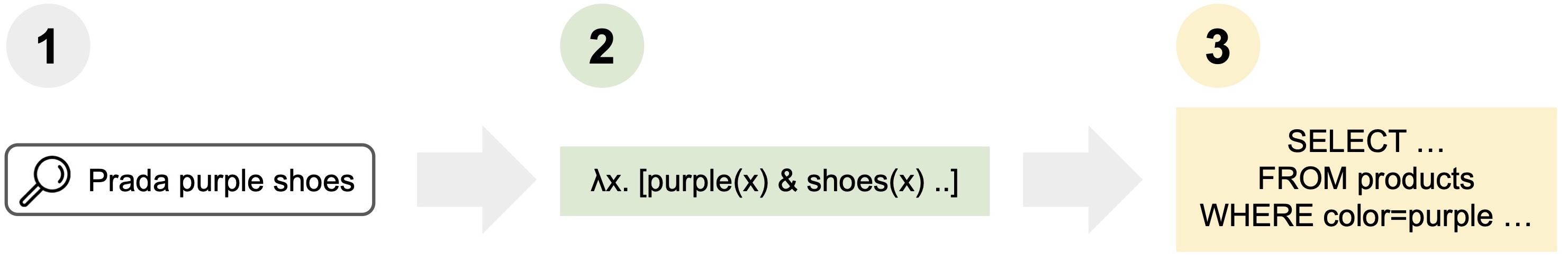}
  \caption{\textbf{From user query (1) to executable code (3), through semantic parsing (2)}. Under the assumption that PO exists, queries could be executed as programmatic instructions through their logical form.}
  \label{fig:search}
  \vspace{-4mm}
\end{figure}

Under the proposed approach, a query is parsed into a logical form (\textit{parsing}), which is mapped to a machine code to be executed over the target domain (\textit{synthesis}): in Fig.~\ref{fig:search} we find lambda expressions and SQL \cite{Hui2021ImprovingTW}, but the proposal is broadly compatible with any explicit formalism. In other words, the \textit{meaning} of ``Prada purple shoes'' is neither boolean operators over TF-IDF weights, nor a BERT-based embedding, but (something like):

$\lambda x. [Purple(x) \;\&\;Shoes(x)\;\&\;Prada(x)]$.\footnote{With $Purple$, $Shoes$, $Prada$ as predicates of type $Color$, $Sortal$, $Brand$.}

Viewing queries as (small) programs to execute has several advantages. First, it provides the ability to apply filters that are already available and show their application to the user -- this is often desirable but in a VSM-system it requires an additional module to be trained and maintained. Second, the explicit and easy-to-debug ``trace'' of the query enables principled fallback strategies. As an illustration, assume the user issues the query ``purple shoes'', which has no perfect matches. The logical form that (roughly) states ``retrieve an object of type shoes, with purple as a color'', allows us to reason about the next best thing available, and provide a graceful fallback message (e.g. ``we don’t have purple shoes, but we thought you could like dark red shoes''\footnote{Note that while IR explanations are often used to improve recommender systems \cite{Zhang2018ExplainableRA}, search may benefit from them for similar reasons.}. An explicit parse leads us to recognize that different tokens in the query have different psychological importance for the shopper: if the retrieval goal revolves around shoes, the system should retrieve items that are still shoes while never retrieving purple items that are not shoes. In this perspective, parsing both yields the exact linguistic intent and lays down possible compositional fallback strategies. Crucially, fallback strategies can be ML-driven, domain-driven, or heuristic-driven and may change from one deployment to the next: by turning queries into code, we make it easier to incorporate constraints (including probabilistic ones) into an interpretable search plan. 

To further appreciate the experience, it is useful to contrast what would have happened under plausible implementations of VSM. Under a sparse vector space, the shopper would typically either get a \textit{No result page}, or -- as the opposite extreme -- received irrelevant items from an OR expansion: non-shoes that are purple, shoes that are green \footnote{Far from being a theoretical possibility, this is the default experience for all website using out open source tools like Elasticsearch, or non-AI SaaS providers.}. Under a dense vector space, retrieval would provide a set of items, but no principled way to cut the set at the right position (when a ``near'' vector is not near enough?) or explain its choice. Both are open problems \cite{10.1145/3397271.3401188}, and no solution is known, especially given data constraints of the RSc \cite{tagliabue2023reasonable}.

There is another, subtler, way to appreciate the impact of PO on search especially relevant for SaaS players \cite{jacopo_medium}, whose job is to develop solutions deployed on dozens independent shops in several languages and verticals. When you have two shops in the same market (as \textbf{Shop A} and \textbf{Shop B} below), PO gets you re-usable abstractions. Overlapping parse trees and product properties can help with cold start scenarios: if a model is matching ``Adidas'' and ``Nike'' as brands with high affinity, it can be ported to a new shop to boostrap learning (i.e. bootstrapping a new ecommerce without any behavioral data). As an even more extreme form of bootstrapping, learning can be transferred between (similar) languages when appropriate resources exist: while VSM models can make good use of multi-language embeddings, the power-law of RSc helps us here as well, as most retailers would do most business in 1-3 languages. \footnote{Even the fallback strategies mentioned before can be ported: if ``sneakers'' is fallback for ``shoes'', the same strategy can be applied any time you have ``shoes'' available in the parse tree.} Of course, re-imagining search with PO opens up possibilities also outside of the search experience itself: just to mention two obvious ones, finer-grained analytics (both about queries as expressing shoppers' intent, and products, as a collection of human-readable properties), and cross-pollination with data coming in and out of the PIM (Product Information Management).

In this section, we argued that a large portion of the market would benefit from program synthesis through semantic parsing, if such a system existed. We now show how such system can be built.

\section{Building a Semantic Parser}
\label{sec:parsing}
As PO itself has two components -- a query and a product parser with a shared domain and interpretation (in the sense of model theory~\cite{sep-model-theory}) -- how do we bootstrap and scale both? Assuming we use ML to train the parser, the hardest part is obtaining a training set for queries: while (almost) any untrained human can annotate an ecommerce catalog, producing logical forms requires a good deal of work by trained linguists. We will therefore break the problem into pieces, by first assuming we have product representations available to build a training set for query parsing, and then relaxing this assumption. 

Fig.~\ref{fig:teaser} showcases the creation of a query dataset for a statistical parser (3) \footnote{The details of the parser are pretty unimportant, as there is substantial evidence that this is a solvable problem with good enough data \cite{wang-etal-2015-building}.}, starting from product representations (1) and a small grammar (2): our insight is that, instead of manual annotation, we can programmatically generate golden triples \textit{<query,logical form,SKUs>} by synthesizing \textit{jointly} queries, their logical form, and the result set, leveraging the isomorphism between product representations and logical forms. Moving the annotation problem away from logical form helps us leverage further insights on the peculiarities of the RSc. First, it should be stressed that extracting (most) product features (1, in Fig.~\ref{fig:teaser}) is easy: some attributes come already structured, and statistically accurate labels are easy to obtain thanks to methods applicable across shops \cite{Gupta2016ProductCI,Naturearticle}. In particular, while recent large language models cannot be directly used at runtime \cite{Ji2022SurveyOH,Liu2023EvaluatingVI}, they are ideally suited to be a complementary strategy to more traditional methods when it comes to entity extraction (or even as an oracle for offline usage \cite{Drozdov2022CompositionalSP}, see the Appendix). Product information is also important for other parts of the business, which means labeling can piggyback on independently motivated processes (e.g. PIM).\footnote{Product labeling can also be outsourced with no privacy concerns.}

Second, the peculiarities of query distribution simplify the slot filling component (2, in Fig.~\ref{fig:teaser}): even in a SaaS scenario where extreme scalability is paramount, NP queries are easy to generate and then re-use  -- the queries ``ski trousers'', ``running shoes'' and ``ski gloves'' (mentioned in \cite{10.1145/3366424.3386198}) share the same logical form. Not only the grammar is simple enough to start, but since the final goal is to parse queries through a model trained on these synthetic NPs, we can err on the side of recall and over-generate (as it will just create training sentences that nobody would use). 

Let's now recap our approach as an actionable list:

\begin{enumerate}
    \item at indexing time, extract product representations from the catalog to be indexed in a knowledge base, through heuristics and/or models ~\cite{Ratner2017SnorkelRT,Naturearticle};\footnote{We refer the readers to the Appendix for more details.}
    \item build a simple NP-focused grammar, to cover a significant part of the distribution. The process can begin by annotating historical queries with simple logical forms, and then generalize a grammar to simplify those trees. To give a sense of how this would work, we selected \textbf{Shop A} and \textbf{Shop B}, multi-brand retailers in the apparel industry and catalog size between 10k and 30k SKUs. We manually annotate historical queries to get a sense of what grammar captures user behavior. Few hundreds parses (respectively, 475 and 459) cover 43\% and 25\% of the entire query distribution for \textbf{Shop A} and \textbf{Shop B};
    \item use the product representation and the NP-grammar to generate a training test with synthetic queries and golden parse trees ( Fig.~\ref{fig:teaser}) -- note that it is easy to augment the set of parsable queries through paraphrases \cite{wang-etal-2015-building} or prompting \cite{Rosenbaum2022};
    \item train a standard parsing model \cite{10.5555/645530.655813,zheng-etal-2017-joint} on this dataset;
    \item at runtime, use the parsing model on an incoming query, get the logical form and map it to an executable code for the target knowledge base: retrieve the products, execute fallback strategies if relevant.
\end{enumerate}

This strategy has consequences for two important pieces of the search experience, \textit{re-ranking} and \textit{type-ahead suggestions}. Re-ranking in VSM is often needed to hide poor results, and may even conflict with relevance objectives: e.g., popular products may sometimes outrank others irrespective of query intent. A structured approach to retrieval allows ranking to be mostly about personalization: given a relevant result set, which of the following ``purple shoes'' is best for this shopper (based on several real-time and historical ranking signals)? Conversely, ranking rules -- both manual and learned -- can be applied on a \textit{ceteris paribus} level: only if two items are equally relevant, popularity can influence their ranking. Query suggestions are known to be important for a good search UX \cite{tagliabue-etal-2020-grow}: synthetic queries (Fig.~\ref{fig:teaser}) could be used to suggest new and cold query types, as well as familiarize shoppers with the capability of the parser; for example, suggesting ``blue shoes under 100 USD'' would gradually educate users in using the search bar better.

\section{Limitations and answers to common concerns}

\subsection{Vectors strike back}
The explosion of NLP-capabilities in recent years have established beyond any reasonable doubt the virtues of distributional semantics \cite{Lake2020WordMI}: it may therefore seem strange to defend program synthesis for IR use cases. The quality of the vectorized representations for queries and products increased dramatically (including exciting possibilities such as multi-modal understanding \cite{chia-etal-2022-come}), but the problem with VLM is still present even in the most sophisticated retailers: as we observe in the result set in Fig.~\ref{fig:amazon}, the query ``nintendo switch'' is retrieving \textit{pens}. While it would be tempting to dismiss this as an artefact or an anecdote, it is on the contrary an essential component of VSM: if relevance is distance in a vector space, there is no \textit{cut-off} establishing when far is \textit{too} far. If we compare the result set to the typical response we would get from a human assistant\footnote{Anecdotally, note that \textit{ChatGPT} response to the prompt ``You are a shopper assistant at Best Buy, the famous electronic retailer. You work in the video-game section. A shopper comes to you and ask for \textit{nintendo switch}: what product do you think she wants to buy?'' is ``If a shopper comes asking for a Nintendo Switch, it's most likely that they are referring to the Nintendo Switch console itself''.}, it is clear the shared meaning of ``nintendo switch'' is very different. For almost-web-scale catalogs, vector search is pragmatically an effective strategy, as the ``very close'' products for most queries are enough to fill the first few result pages; for smaller catalogs, however, the perceived relevance may quickly degrade and the VSM approach has no principled countermeasure. 

As we discuss below, better vector representations are an essential component of \textit{any} search engine, and NLP breakthroughs are a welcome addition to the toolkit of any shop. However, treating relevance solely as a distance calculation is an approximation, and should be recognized as such: when we switch our attention from lexically-driven to compositionally-driven use cases, how much value can we now unlock? 

\subsection{Parsing vs rewriting}

Parsing is hardly the only query processing technique available to RSc shops: for example, query rewriting is a popular approach to bridge the gap between the user's intent (``red Nike sneakers'') and inventory (burgundy Adidas shoes). However, it is important to realize that the concerns of parsing and rewriting modules are distinct, and possibly complementary: you can rewrite ``sneakers'' into ``shoes'' before parsing it into an object type, but rewriting by itself does not challenge the fundamental assumption of VSM -- effective rewriting may improve recall, but does not unlock any of the relevance benefit that parsing provides (Section~\ref{sec:oracle}). From an engineering perspective, it's easy to see how a rewriting module could leave completely untouched the retrieval machinery of VSM, while parsing requires re-thinking the strategy entirely. 

Moreover, a crucial component of our proposal is the ``zero-shot'' adaptation obtained through the loose isomorphism between products in a graph and grammars: since parsing is built through product understanding, not explicit or implicit behavioral supervision, its sample efficiency makes it ideal for RSc shops and horizontal scalability (see below); on the other side, modern NLP-based rewriting through behavioral supervision \cite{Wang2021} is better suited for big retailers.\footnote{Five years after the deployment of the system in \textit{this} paper, it is telling that leading tech retailers are starting to use a product graph for rewriting as well \cite{10.1145/3543873.3587678}.}

\subsection{Vertical vs horizontal scaling}
When thinking about ``scalable'' engineering, we think of diminishing marginal effort as we ``scale'' along an important dimension. Since most IR is done at Big Tech scale, the implicit notion of scalability is the B2C one: as a target shop grows in inventory and traffic, the long tail of queries will expand and rare events become more important (Section~\ref{sec:industry}). In this regime, data-driven approaches are \textit{scalable}: the more traffic, the more data, so statistical generalization is a promising path to diminishing marginal effort -- how hard is to satisfy this shopper's intent, given we have seen already \textit{k} million of them? 

As we hinted in \textit{this} work, there is another concept of scalability in IR, which becomes evident for B2B scenarios: if our system is used across multiple RSc shops, the marginal cost that will dominate the business is \textit{deployment} cost -- how hard is to get a new shop online, given we put online \textit{k} already? The synthesis approach we championed has been developed mainly targeting this second notion: if the marginal cost of tagging catalogs is diminishing (see the \textit{Appendix}), the cost of understanding queries on newer shops diminishes as well, \textit{irrespective} of how much traffic they get. While emphasis has been put on synthesis as the actual implementation mechanism for our strategy, the broader, and perhaps novel insight, is that query performance is (in certain cases) a by-product of product understanding and linguistic knowledge, both of which are more scalable than practitioners typically realize.

\subsection{Parser fragility}

A critical point that has not been addressed is the ``fragility'' of parsing-first strategies: since no parsing model would be perfect, what should we do when it fails? In our experience, the most natural architecture is a two-tier system, such that, if parsing or program execution fail, the system would resort to a traditional VSM strategy (e.g. a sparse / dense vector-based retrieval). Considering the speed of an ML parser, we pay a tiny latency tax for the above mentioned benefits. When it comes to deployment, our recommendation is to use program synthesis on top of a basic VSM retrieval, not as a replacement; philosophically however, our position remains that VSM is an \textit{approximation} to relevance, and should be treated as such.

\section{Conclusion}

Motivated by query distributions and industry constraints, we argued that program synthesis (through semantic parsing) is a feasible path for a better search experience at RSc, compared to VSM alone as a relevance model. We showed that the usual worries associated with explicit meaning representations are unwarranted, and maintained that the key insight to a novel view on search is the ``isomorphic'' structure of (parsed) queries and product structure.

The representation dichotomy \textit{explicit-but-annotation-heavy} VSM \textit{approximate-but-fully-learnable} is indeed a false one, and we sketched how a tiny initial linguistic structure can help bootstrapping a large-scale parsing system.  We are confident through 6 years of experience, deployments and publications that RSc shops can benefit from it, and we hope \textit{this} paper will start a discussion with participants coming from different backgrounds. While this work hardly constitutes the last word on the topic, it is hopefully a first step in leading the field away from local optima, and embracing the peculiarities and opportunities of product search.

\begin{acks}
The dry prose of a scholarly paper cannot do justice to the adventure that is building an early-stage startup: this paper would not have been possible without \textit{Tooso}, the company pioneering search-as-parsing at scale back in 2018-2019. We wish to thank first Mattia, Luca, Andrea, Alessia, and then everybody else involved in that clumsy, special company: a challenge we were willing to accept, one we were unwilling to postpone, and one we intended to win.

Furthermore, we wish to thank Tracy Holloway King, Federico Bianchi, Patrick John Chia and two anonymous reviewers for useful comments to a previous version of this paper.
\end{acks}

\bibliographystyle{ACM-Reference-Format}
\bibliography{sample-base}


\begin{thebibliography}{58}


\ifx \showCODEN    \undefined \def \showCODEN     #1{\unskip}     \fi
\ifx \showDOI      \undefined \def \showDOI       #1{#1}\fi
\ifx \showISBNx    \undefined \def \showISBNx     #1{\unskip}     \fi
\ifx \showISBNxiii \undefined \def \showISBNxiii  #1{\unskip}     \fi
\ifx \showISSN     \undefined \def \showISSN      #1{\unskip}     \fi
\ifx \showLCCN     \undefined \def \showLCCN      #1{\unskip}     \fi
\ifx \shownote     \undefined \def \shownote      #1{#1}          \fi
\ifx \showarticletitle \undefined \def \showarticletitle #1{#1}   \fi
\ifx \showURL      \undefined \def \showURL       {\relax}        \fi
\providecommand\bibfield[2]{#2}
\providecommand\bibinfo[2]{#2}
\providecommand\natexlab[1]{#1}
\providecommand\showeprint[2][]{arXiv:#2}

\bibitem[Ai and Narayanan.R(2021)]%
        {10.1145/3459637.3482276}
\bibfield{author}{\bibinfo{person}{Qingyao Ai} {and} \bibinfo{person}{Lakshmi
  Narayanan.R}.} \bibinfo{year}{2021}\natexlab{}.
\newblock \showarticletitle{Model-Agnostic vs. Model-Intrinsic Interpretability
  for Explainable Product Search}. In \bibinfo{booktitle}{\emph{Proceedings of
  the 30th ACM International Conference on Information and Knowledge
  Management}} (Virtual Event, Queensland, Australia)
  \emph{(\bibinfo{series}{CIKM '21})}. \bibinfo{publisher}{Association for
  Computing Machinery}, \bibinfo{address}{New York, NY, USA},
  \bibinfo{pages}{5–15}.
\newblock
\showISBNx{9781450384469}
\urldef\tempurl%
\url{https://doi.org/10.1145/3459637.3482276}
\showDOI{\tempurl}


\bibitem[Alaimo(2018)]%
        {retailDive2020}
\bibfield{author}{\bibinfo{person}{Dan Alaimo}.}
  \bibinfo{year}{2018}\natexlab{}.
\newblock \bibinfo{booktitle}{\emph{87\% of shoppers now begin product searches
  online.}}
\newblock
\urldef\tempurl%
\url{https://www.retaildive.com/news/87-of-shoppers-now-begin-product-searches-online/530139/}
\showURL{%
\tempurl}


\bibitem[Bahri et~al\mbox{.}(2020)]%
        {10.1145/3397271.3401188}
\bibfield{author}{\bibinfo{person}{Dara Bahri}, \bibinfo{person}{Yi Tay},
  \bibinfo{person}{Che Zheng}, \bibinfo{person}{Donald Metzler}, {and}
  \bibinfo{person}{Andrew Tomkins}.} \bibinfo{year}{2020}\natexlab{}.
\newblock \showarticletitle{Choppy: Cut Transformer for Ranked List
  Truncation}. In \bibinfo{booktitle}{\emph{Proceedings of the 43rd
  International ACM SIGIR Conference on Research and Development in Information
  Retrieval}} (Virtual Event, China) \emph{(\bibinfo{series}{SIGIR '20})}.
  \bibinfo{publisher}{Association for Computing Machinery},
  \bibinfo{address}{New York, NY, USA}, \bibinfo{pages}{1513–1516}.
\newblock
\showISBNx{9781450380164}
\urldef\tempurl%
\url{https://doi.org/10.1145/3397271.3401188}
\showDOI{\tempurl}


\bibitem[Basin et~al\mbox{.}(2004)]%
        {inproceedings}
\bibfield{author}{\bibinfo{person}{David Basin}, \bibinfo{person}{Yves
  Deville}, \bibinfo{person}{Pierre Flener}, \bibinfo{person}{Andreas Hamfelt},
  {and} \bibinfo{person}{Jørgen Nilsson}.} \bibinfo{year}{2004}\natexlab{}.
\newblock \showarticletitle{Synthesis of Programs in Computational Logic}.
\newblock \bibinfo{journal}{\emph{Program Development in Computational Logic}}
  \bibinfo{volume}{3049}, \bibinfo{pages}{30--65}.
\newblock
\showISBNx{978-3-540-22152-4}
\urldef\tempurl%
\url{https://doi.org/10.1007/978-3-540-25951-0_2}
\showDOI{\tempurl}


\bibitem[Berg et~al\mbox{.}(2019)]%
        {Metaflow}
\bibfield{author}{\bibinfo{person}{David Berg}, \bibinfo{person}{Ravi~Kiran
  Chirravuri}, \bibinfo{person}{Romain Cledat}, \bibinfo{person}{Savin Goyal},
  \bibinfo{person}{Ferras Hamad}, {and} \bibinfo{person}{Ville Tuulos}.}
  \bibinfo{year}{2019}\natexlab{}.
\newblock \bibinfo{booktitle}{\emph{Open-Sourcing Metaflow, a Human-Centric
  Framework for Data Science}}.
\newblock
\urldef\tempurl%
\url{https://netflixtechblog.com/open-sourcing-metaflow-a-human-centric-framework-for-data-science-fa72e04a5d9}
\showURL{%
\tempurl}


\bibitem[Bianchi et~al\mbox{.}(2021a)]%
        {bianchi-etal-2021-language}
\bibfield{author}{\bibinfo{person}{Federico Bianchi}, \bibinfo{person}{Ciro
  Greco}, {and} \bibinfo{person}{Jacopo Tagliabue}.}
  \bibinfo{year}{2021}\natexlab{a}.
\newblock \showarticletitle{Language in a (Search) Box: Grounding Language
  Learning in Real-World Human-Machine Interaction}. In
  \bibinfo{booktitle}{\emph{Proceedings of the 2021 Conference of the North
  American Chapter of the Association for Computational Linguistics: Human
  Language Technologies}}. \bibinfo{publisher}{Association for Computational
  Linguistics}, \bibinfo{address}{Online}, \bibinfo{pages}{4409--4415}.
\newblock
\urldef\tempurl%
\url{https://doi.org/10.18653/v1/2021.naacl-main.348}
\showDOI{\tempurl}


\bibitem[Bianchi et~al\mbox{.}(2021b)]%
        {bianchi-etal-2021-query2prod2vec}
\bibfield{author}{\bibinfo{person}{Federico Bianchi}, \bibinfo{person}{Jacopo
  Tagliabue}, {and} \bibinfo{person}{Bingqing Yu}.}
  \bibinfo{year}{2021}\natexlab{b}.
\newblock \showarticletitle{{Q}uery2{P}rod2{V}ec: Grounded Word Embeddings for
  e{C}ommerce}. In \bibinfo{booktitle}{\emph{Proceedings of the 2021 Conference
  of the North American Chapter of the Association for Computational
  Linguistics: Human Language Technologies: Industry Papers}}.
  \bibinfo{publisher}{Association for Computational Linguistics},
  \bibinfo{address}{Online}, \bibinfo{pages}{154--162}.
\newblock
\urldef\tempurl%
\url{https://doi.org/10.18653/v1/2021.naacl-industry.20}
\showDOI{\tempurl}


\bibitem[Bianchi et~al\mbox{.}(2020)]%
        {BianchiSIGIReCom2020}
\bibfield{author}{\bibinfo{person}{Federico Bianchi}, \bibinfo{person}{Jacopo
  Tagliabue}, \bibinfo{person}{Bingqing Yu}, \bibinfo{person}{Luca Bigon},
  {and} \bibinfo{person}{Ciro Greco}.} \bibinfo{year}{2020}\natexlab{}.
\newblock \showarticletitle{Fantastic Embeddings and How to Align Them:
  Zero-Shot Inference in a Multi-Shop Scenario}. In
  \bibinfo{booktitle}{\emph{Proceedings of the SIGIR 2020 eCom workshop, July
  2020, Virtual Event, published at http://ceur-ws.org (to appear)}}.
\newblock
\urldef\tempurl%
\url{https://arxiv.org/abs/2007.14906}
\showURL{%
\tempurl}


\bibitem[{Bloomreach}(2022)]%
        {BloomreachRound}
\bibfield{author}{\bibinfo{person}{{Bloomreach}}.}
  \bibinfo{year}{2022}\natexlab{}.
\newblock \bibinfo{booktitle}{\emph{With \$175 Million in Funding, Bloomreach
  Is Authoring the Next Chapter of E-Commerce}}.
\newblock
\urldef\tempurl%
\url{https://www.bloomreach.com/en/blog/2022/with-usd175-million-in-funding-bloomreach-is-authoring-the-next-chapter-of-e-commerce}
\showURL{%
\tempurl}


\bibitem[Brenner et~al\mbox{.}(2018)]%
        {Brenner2018EndtoEndNR}
\bibfield{author}{\bibinfo{person}{Eliot Brenner}, \bibinfo{person}{Jun Zhao},
  \bibinfo{person}{Aliasgar Kutiyanawala}, {and} \bibinfo{person}{Zheng Yan}.}
  \bibinfo{year}{2018}\natexlab{}.
\newblock \showarticletitle{End-to-End Neural Ranking for eCommerce Product
  Search: an Application of Task Models and Textual Embeddings}.
\newblock \bibinfo{journal}{\emph{ArXiv}}  \bibinfo{volume}{abs/1806.07296}
  (\bibinfo{year}{2018}).
\newblock


\bibitem[Chia et~al\mbox{.}(2022a)]%
        {Naturearticle}
\bibfield{author}{\bibinfo{person}{Patrick Chia}, \bibinfo{person}{Giuseppe
  Attanasio}, \bibinfo{person}{Federico Bianchi}, \bibinfo{person}{Silvia
  Terragni}, \bibinfo{person}{Ana Magalhães}, \bibinfo{person}{Diogo
  Goncalves}, \bibinfo{person}{Ciro Greco}, {and} \bibinfo{person}{Jacopo
  Tagliabue}.} \bibinfo{year}{2022}\natexlab{a}.
\newblock \showarticletitle{Contrastive language and vision learning of general
  fashion concepts}.
\newblock \bibinfo{journal}{\emph{Scientific Reports}}  \bibinfo{volume}{12}
  (\bibinfo{date}{11} \bibinfo{year}{2022}).
\newblock
\urldef\tempurl%
\url{https://doi.org/10.1038/s41598-022-23052-9}
\showDOI{\tempurl}


\bibitem[Chia et~al\mbox{.}(2022b)]%
        {chia-etal-2022-come}
\bibfield{author}{\bibinfo{person}{Patrick~John Chia}, \bibinfo{person}{Jacopo
  Tagliabue}, \bibinfo{person}{Federico Bianchi}, \bibinfo{person}{Ciro Greco},
  {and} \bibinfo{person}{Diogo Goncalves}.} \bibinfo{year}{2022}\natexlab{b}.
\newblock \showarticletitle{{``}Does it come in black?{''} {CLIP}-like models
  are zero-shot recommenders}. In \bibinfo{booktitle}{\emph{Proceedings of the
  Fifth Workshop on e-Commerce and NLP (ECNLP 5)}}.
  \bibinfo{publisher}{Association for Computational Linguistics},
  \bibinfo{address}{Dublin, Ireland}, \bibinfo{pages}{191--198}.
\newblock
\urldef\tempurl%
\url{https://doi.org/10.18653/v1/2022.ecnlp-1.22}
\showDOI{\tempurl}


\bibitem[Choudhary et~al\mbox{.}(2022)]%
        {choudhary2022anthem}
\bibfield{author}{\bibinfo{person}{Nurendra Choudhary}, \bibinfo{person}{Nikhil
  Rao}, \bibinfo{person}{Sumeet Katariya}, \bibinfo{person}{Karthik Subbian},
  {and} \bibinfo{person}{Chandan~K. Reddy}.} \bibinfo{year}{2022}\natexlab{}.
\newblock \showarticletitle{ANTHEM: Attentive Hyperbolic Entity Model for
  Product Search}. In \bibinfo{booktitle}{\emph{{WSDM} '22: The Fifteenth {ACM}
  International Conference on Web Search and Data Mining, Phoenix, AZ, USA,
  February 21-25, 2022}} (Phoenix, AZ, USA) \emph{(\bibinfo{series}{WSDM
  '22})}. \bibinfo{publisher}{Association for Computing Machinery},
  \bibinfo{address}{New York, NY, USA}.
\newblock


\bibitem[Commerce(2021)]%
        {bigCommerce2020}
\bibfield{author}{\bibinfo{person}{Big Commerce}.}
  \bibinfo{year}{2021}\natexlab{}.
\newblock \bibinfo{booktitle}{\emph{How Ecommerce Site Search Can Create a
  Competitive Advantage.}}
\newblock
\urldef\tempurl%
\url{https://www.bigcommerce.com/articles/ecommerce/site-search/#the-effectiveness-of-ecommerce-site-search-}
\showURL{%
\tempurl}


\bibitem[Compton(2021)]%
        {forrester2021}
\bibfield{author}{\bibinfo{person}{Scott Compton}.}
  \bibinfo{year}{2021}\natexlab{}.
\newblock \bibinfo{booktitle}{\emph{Searching For ROI In Retail: The Time For A
  New Site Search Tool Is Now}}.
\newblock
\urldef\tempurl%
\url{https://www.forrester.com/blogs/searching-for-roi-in-retail-the-time-for-a-new-site-search-tool-is-now/?categoryid=a89c0000000AKp1AAG}
\showURL{%
\tempurl}


\bibitem[Cramer-Flood(2020)]%
        {emarketer2020}
\bibfield{author}{\bibinfo{person}{Ethan Cramer-Flood}.}
  \bibinfo{year}{2020}\natexlab{}.
\newblock \bibinfo{booktitle}{\emph{Global Ecommerce 2020. Ecommerce
  Decelerates amid Global Retail Contraction but Remains a Bright Spot.}}
\newblock
\urldef\tempurl%
\url{https://www.emarketer.com/content/global-ecommerce-2020}
\showURL{%
\tempurl}


\bibitem[Drozdov et~al\mbox{.}(2022)]%
        {Drozdov2022CompositionalSP}
\bibfield{author}{\bibinfo{person}{Andrew Drozdov}, \bibinfo{person}{Nathanael
  Scharli}, \bibinfo{person}{Ekin Akyuurek}, \bibinfo{person}{Nathan Scales},
  \bibinfo{person}{Xinying Song}, \bibinfo{person}{Xinyun Chen},
  \bibinfo{person}{Olivier Bousquet}, {and} \bibinfo{person}{Denny Zhou}.}
  \bibinfo{year}{2022}\natexlab{}.
\newblock \showarticletitle{Compositional Semantic Parsing with Large Language
  Models}.
\newblock \bibinfo{journal}{\emph{ArXiv}}  \bibinfo{volume}{abs/2209.15003}
  (\bibinfo{year}{2022}).
\newblock


\bibitem[Eric(2022)]%
        {ericMLOPSMess}
\bibfield{author}{\bibinfo{person}{Mihail Eric}.}
  \bibinfo{year}{2022}\natexlab{}.
\newblock \bibinfo{title}{MLOps Is a Mess But That's to be Expected}.
\newblock
\newblock
\urldef\tempurl%
\url{https://www.mihaileric.com/posts/mlops-is-a-mess/}
\showURL{%
\tempurl}


\bibitem[Farzana et~al\mbox{.}(2023)]%
        {10.1145/3543873.3587678}
\bibfield{author}{\bibinfo{person}{Shahla Farzana}, \bibinfo{person}{Qunzhi
  Zhou}, {and} \bibinfo{person}{Petar Ristoski}.}
  \bibinfo{year}{2023}\natexlab{}.
\newblock \showarticletitle{Knowledge Graph-Enhanced Neural Query Rewriting}.
  In \bibinfo{booktitle}{\emph{Companion Proceedings of the ACM Web Conference
  2023}} (Austin, TX, USA) \emph{(\bibinfo{series}{WWW '23 Companion})}.
  \bibinfo{publisher}{Association for Computing Machinery},
  \bibinfo{address}{New York, NY, USA}, \bibinfo{pages}{911–919}.
\newblock
\showISBNx{9781450394192}
\urldef\tempurl%
\url{https://doi.org/10.1145/3543873.3587678}
\showDOI{\tempurl}


\bibitem[{Gillick} et~al\mbox{.}(2018)]%
        {gillick2018end}
\bibfield{author}{\bibinfo{person}{Daniel {Gillick}},
  \bibinfo{person}{Alessandro {Presta}}, {and} \bibinfo{person}{Gaurav~Singh
  {Tomar}}.} \bibinfo{year}{2018}\natexlab{}.
\newblock \showarticletitle{End-to-End Retrieval in Continuous Space}.
\newblock \bibinfo{journal}{\emph{arXiv preprint arXiv:1811.08008}}
  (\bibinfo{year}{2018}).
\newblock


\bibitem[Gupta et~al\mbox{.}(2016)]%
        {Gupta2016ProductCI}
\bibfield{author}{\bibinfo{person}{Vivek Gupta}, \bibinfo{person}{Harish
  Karnick}, \bibinfo{person}{Ashendra Bansal}, {and} \bibinfo{person}{Pradhuman
  Jhala}.} \bibinfo{year}{2016}\natexlab{}.
\newblock \showarticletitle{Product Classification in E-Commerce using
  Distributional Semantics}. In \bibinfo{booktitle}{\emph{COLING}}.
\newblock


\bibitem[Hodges(2022)]%
        {sep-model-theory}
\bibfield{author}{\bibinfo{person}{Wilfrid Hodges}.}
  \bibinfo{year}{2022}\natexlab{}.
\newblock \showarticletitle{{Model Theory}}.
\newblock In \bibinfo{booktitle}{\emph{The {Stanford} Encyclopedia of
  Philosophy} (\bibinfo{edition}{{S}pring 2022} ed.)},
  \bibfield{editor}{\bibinfo{person}{Edward~N. Zalta}} (Ed.).
  \bibinfo{publisher}{Metaphysics Research Lab, Stanford University}.
\newblock


\bibitem[Hui et~al\mbox{.}(2021)]%
        {Hui2021ImprovingTW}
\bibfield{author}{\bibinfo{person}{Binyuan Hui}, \bibinfo{person}{Xiang Shi},
  \bibinfo{person}{Ruiying Geng}, \bibinfo{person}{Binhua Li},
  \bibinfo{person}{Yongbin Li}, \bibinfo{person}{Jian Sun}, {and}
  \bibinfo{person}{Xiaodan Zhu}.} \bibinfo{year}{2021}\natexlab{}.
\newblock \showarticletitle{Improving Text-to-SQL with Schema Dependency
  Learning}.
\newblock \bibinfo{journal}{\emph{ArXiv}}  \bibinfo{volume}{abs/2103.04399}
  (\bibinfo{year}{2021}).
\newblock


\bibitem[Ji et~al\mbox{.}(2022)]%
        {Ji2022SurveyOH}
\bibfield{author}{\bibinfo{person}{Ziwei Ji}, \bibinfo{person}{Nayeon Lee},
  \bibinfo{person}{Rita Frieske}, \bibinfo{person}{Tiezheng Yu},
  \bibinfo{person}{Dan Su}, \bibinfo{person}{Yan Xu}, \bibinfo{person}{Etsuko
  Ishii}, \bibinfo{person}{Yejin Bang}, \bibinfo{person}{Wenliang Dai},
  \bibinfo{person}{Andrea Madotto}, {and} \bibinfo{person}{Pascale Fung}.}
  \bibinfo{year}{2022}\natexlab{}.
\newblock \showarticletitle{Survey of Hallucination in Natural Language
  Generation}.
\newblock \bibinfo{journal}{\emph{Comput. Surveys}}  \bibinfo{volume}{55}
  (\bibinfo{year}{2022}), \bibinfo{pages}{1 -- 38}.
\newblock


\bibitem[Jia and Liang(2016)]%
        {Jia2016DataRF}
\bibfield{author}{\bibinfo{person}{Robin Jia} {and} \bibinfo{person}{Percy
  Liang}.} \bibinfo{year}{2016}\natexlab{}.
\newblock \showarticletitle{Data Recombination for Neural Semantic Parsing}.
\newblock \bibinfo{journal}{\emph{ArXiv}}  \bibinfo{volume}{abs/1606.03622}
  (\bibinfo{year}{2016}).
\newblock


\bibitem[Karpukhin et~al\mbox{.}(2020)]%
        {Karpukhin2020DensePR}
\bibfield{author}{\bibinfo{person}{Vladimir Karpukhin}, \bibinfo{person}{Barlas
  Oğuz}, \bibinfo{person}{Sewon Min}, \bibinfo{person}{Patrick Lewis},
  \bibinfo{person}{Ledell~Yu Wu}, \bibinfo{person}{Sergey Edunov},
  \bibinfo{person}{Danqi Chen}, {and} \bibinfo{person}{Wen tau Yih}.}
  \bibinfo{year}{2020}\natexlab{}.
\newblock \showarticletitle{Dense Passage Retrieval for Open-Domain Question
  Answering}.
\newblock \bibinfo{journal}{\emph{ArXiv}}  \bibinfo{volume}{abs/2004.04906}
  (\bibinfo{year}{2020}).
\newblock


\bibitem[King(2023)]%
        {King2023}
\bibfield{author}{\bibinfo{person}{Tracy~Holloway King}.}
  \bibinfo{year}{2023}\natexlab{}.
\newblock \bibinfo{booktitle}{\emph{White Roses, Red Backgrounds: Bringing
  Structured Representations to Search}}.
\newblock \bibinfo{publisher}{Springer International Publishing},
  \bibinfo{address}{Cham}, \bibinfo{pages}{191--215}.
\newblock
\showISBNx{978-3-031-21780-7}
\urldef\tempurl%
\url{https://doi.org/10.1007/978-3-031-21780-7_9}
\showDOI{\tempurl}


\bibitem[Lafferty et~al\mbox{.}(2001)]%
        {10.5555/645530.655813}
\bibfield{author}{\bibinfo{person}{John~D. Lafferty}, \bibinfo{person}{Andrew
  McCallum}, {and} \bibinfo{person}{Fernando C.~N. Pereira}.}
  \bibinfo{year}{2001}\natexlab{}.
\newblock \showarticletitle{Conditional Random Fields: Probabilistic Models for
  Segmenting and Labeling Sequence Data}. In
  \bibinfo{booktitle}{\emph{Proceedings of the Eighteenth International
  Conference on Machine Learning}} \emph{(\bibinfo{series}{ICML '01})}.
  \bibinfo{publisher}{Morgan Kaufmann Publishers Inc.}, \bibinfo{address}{San
  Francisco, CA, USA}, \bibinfo{pages}{282–289}.
\newblock
\showISBNx{1558607781}


\bibitem[Lake and Murphy(2020)]%
        {Lake2020WordMI}
\bibfield{author}{\bibinfo{person}{Brenden~M. Lake} {and}
  \bibinfo{person}{Gregory~L. Murphy}.} \bibinfo{year}{2020}\natexlab{}.
\newblock \showarticletitle{Word meaning in minds and machines}.
\newblock \bibinfo{journal}{\emph{Psychological review}}
  (\bibinfo{year}{2020}).
\newblock


\bibitem[Li et~al\mbox{.}(2019)]%
        {Li2019FromSR}
\bibfield{author}{\bibinfo{person}{Rui Li}, \bibinfo{person}{Yunjiang Jiang},
  \bibinfo{person}{Wenyun Yang}, \bibinfo{person}{Guoyu Tang},
  \bibinfo{person}{Songlin Wang}, \bibinfo{person}{Chaoyi Ma},
  \bibinfo{person}{Wei He}, \bibinfo{person}{Xi Xiong}, \bibinfo{person}{Yun
  Xiao}, {and} \bibinfo{person}{Yihong~Eric Zhao}.}
  \bibinfo{year}{2019}\natexlab{}.
\newblock \showarticletitle{From Semantic Retrieval to Pairwise Ranking:
  Applying Deep Learning in E-commerce Search}.
\newblock \bibinfo{journal}{\emph{Proceedings of the 42nd International ACM
  SIGIR Conference on Research and Development in Information Retrieval}}
  (\bibinfo{year}{2019}).
\newblock


\bibitem[Liu et~al\mbox{.}(2023)]%
        {Liu2023EvaluatingVI}
\bibfield{author}{\bibinfo{person}{Nelson~F. Liu}, \bibinfo{person}{Tianyi
  Zhang}, {and} \bibinfo{person}{Percy Liang}.}
  \bibinfo{year}{2023}\natexlab{}.
\newblock \showarticletitle{Evaluating Verifiability in Generative Search
  Engines}.
\newblock


\bibitem[Marotta(2021)]%
        {CoveoRound}
\bibfield{author}{\bibinfo{person}{Stefanie Marotta}.}
  \bibinfo{year}{2021}\natexlab{}.
\newblock \bibinfo{booktitle}{\emph{Canada’s Latest Tech Public Debut Swings
  Amid Soft IPOs}}.
\newblock
\urldef\tempurl%
\url{https://www.bloomberg.com/news/articles/2021-11-25/canada-s-latest-tech-public-debut-swings-amid-slew-of-soft-ipos}
\showURL{%
\tempurl}


\bibitem[Mitra and Craswell(2018)]%
        {mitra2018an}
\bibfield{author}{\bibinfo{person}{Bhaskar Mitra} {and} \bibinfo{person}{Nick
  Craswell}.} \bibinfo{year}{2018}\natexlab{}.
\newblock \showarticletitle{An Introduction to Neural Information Retrieval}.
\newblock \bibinfo{journal}{\emph{Foundations and Trends® in Information
  Retrieval}} \bibinfo{volume}{13}, \bibinfo{number}{1}
  (\bibinfo{date}{December} \bibinfo{year}{2018}), \bibinfo{pages}{1--126}.
\newblock
\urldef\tempurl%
\url{https://www.microsoft.com/en-us/research/publication/introduction-neural-information-retrieval/}
\showURL{%
\tempurl}


\bibitem[Molino and R\'{e}(2021)]%
        {10.1145/3475965.3479315}
\bibfield{author}{\bibinfo{person}{Piero Molino} {and}
  \bibinfo{person}{Christopher R\'{e}}.} \bibinfo{year}{2021}\natexlab{}.
\newblock \showarticletitle{Declarative Machine Learning Systems: The Future of
  Machine Learning Will Depend on It Being in the Hands of the Rest of Us.}
\newblock \bibinfo{journal}{\emph{Queue}} \bibinfo{volume}{19},
  \bibinfo{number}{3} (\bibinfo{date}{jun} \bibinfo{year}{2021}),
  \bibinfo{pages}{46–76}.
\newblock
\showISSN{1542-7730}
\urldef\tempurl%
\url{https://doi.org/10.1145/3475965.3479315}
\showDOI{\tempurl}


\bibitem[Nogueira et~al\mbox{.}(2019)]%
        {Nogueira2019MultiStageDR}
\bibfield{author}{\bibinfo{person}{Rodrigo Nogueira}, \bibinfo{person}{Wei
  Yang}, \bibinfo{person}{Kyunghyun Cho}, {and} \bibinfo{person}{Jimmy~J.
  Lin}.} \bibinfo{year}{2019}\natexlab{}.
\newblock \showarticletitle{Multi-Stage Document Ranking with BERT}.
\newblock \bibinfo{journal}{\emph{ArXiv}}  \bibinfo{volume}{abs/1910.14424}
  (\bibinfo{year}{2019}).
\newblock


\bibitem[Pei et~al\mbox{.}(2019)]%
        {Pei2019PersonalizedCR}
\bibfield{author}{\bibinfo{person}{Changhua Pei}, \bibinfo{person}{Yi Zhang},
  \bibinfo{person}{Yongfeng Zhang}, \bibinfo{person}{Fei Sun},
  \bibinfo{person}{Xiao Lin}, \bibinfo{person}{Hanxiao Sun},
  \bibinfo{person}{Jian Wu}, \bibinfo{person}{Peng Jiang},
  \bibinfo{person}{Wenwu Ou}, {and} \bibinfo{person}{Dan Pei}.}
  \bibinfo{year}{2019}\natexlab{}.
\newblock \showarticletitle{Personalized Context-aware Re-ranking for
  E-commerce Recommender Systems}.
\newblock \bibinfo{journal}{\emph{ArXiv}}  \bibinfo{volume}{abs/1904.06813}
  (\bibinfo{year}{2019}).
\newblock


\bibitem[Ratner et~al\mbox{.}(2017)]%
        {Ratner2017SnorkelRT}
\bibfield{author}{\bibinfo{person}{Alexander~J. Ratner},
  \bibinfo{person}{Stephen~H. Bach}, \bibinfo{person}{Henry~R. Ehrenberg},
  \bibinfo{person}{Jason~Alan Fries}, \bibinfo{person}{Sen Wu}, {and}
  \bibinfo{person}{Christopher R{\'e}}.} \bibinfo{year}{2017}\natexlab{}.
\newblock \showarticletitle{Snorkel: Rapid Training Data Creation with Weak
  Supervision}.
\newblock \bibinfo{journal}{\emph{Proceedings of the VLDB Endowment.
  International Conference on Very Large Data Bases}}  \bibinfo{volume}{11 3}
  (\bibinfo{year}{2017}), \bibinfo{pages}{269--282}.
\newblock


\bibitem[Requena et~al\mbox{.}(2020)]%
        {Requena2020}
\bibfield{author}{\bibinfo{person}{Borja Requena}, \bibinfo{person}{Giovanni
  Cassani}, \bibinfo{person}{Jacopo Tagliabue}, \bibinfo{person}{Ciro Greco},
  {and} \bibinfo{person}{Lucas Lacasa}.} \bibinfo{year}{2020}\natexlab{}.
\newblock \showarticletitle{Shopper intent prediction from clickstream
  e-commerce data with minimal browsing information}.
\newblock \bibinfo{journal}{\emph{Scientific Reports}}  \bibinfo{volume}{10}
  (\bibinfo{year}{2020}), \bibinfo{pages}{2045--2322}.
\newblock
\urldef\tempurl%
\url{https://doi.org/10.1038/s41598-020-73622-y}
\showDOI{\tempurl}


\bibitem[Rosenbaum et~al\mbox{.}(2022)]%
        {Rosenbaum2022}
\bibfield{author}{\bibinfo{person}{Andy Rosenbaum}, \bibinfo{person}{Saleh
  Soltan}, \bibinfo{person}{Wael Hamza}, \bibinfo{person}{Amir Saffari},
  \bibinfo{person}{Marco Damonte}, {and} \bibinfo{person}{Isabel Groves}.}
  \bibinfo{year}{2022}\natexlab{}.
\newblock \showarticletitle{CLASP: Few-shot cross-lingual data augmentation for
  semantic parsing}. In \bibinfo{booktitle}{\emph{AACL-IJCNLP 2022}}.
\newblock
\urldef\tempurl%
\url{https://www.amazon.science/publications/clasp-few-shot-cross-lingual-data-augmentation-for-semantic-parsing}
\showURL{%
\tempurl}


\bibitem[Salton et~al\mbox{.}(1975)]%
        {10.1145/361219.361220}
\bibfield{author}{\bibinfo{person}{G. Salton}, \bibinfo{person}{A. Wong}, {and}
  \bibinfo{person}{C.~S. Yang}.} \bibinfo{year}{1975}\natexlab{}.
\newblock \showarticletitle{A Vector Space Model for Automatic Indexing}.
\newblock \bibinfo{journal}{\emph{Commun. ACM}} \bibinfo{volume}{18},
  \bibinfo{number}{11} (\bibinfo{date}{nov} \bibinfo{year}{1975}),
  \bibinfo{pages}{613–620}.
\newblock
\showISSN{0001-0782}
\urldef\tempurl%
\url{https://doi.org/10.1145/361219.361220}
\showDOI{\tempurl}


\bibitem[Schade and Nielsen(2022)]%
        {userSearch}
\bibfield{author}{\bibinfo{person}{Amy Schade} {and} \bibinfo{person}{Jakob
  Nielsen}.} \bibinfo{year}{2022}\natexlab{}.
\newblock \bibinfo{booktitle}{\emph{Ecommerce User Experience Vol. 05:
  Search.}}
\newblock
\urldef\tempurl%
\url{https://www.nngroup.com/reports/ecommerce-ux-search-including-faceted-search/}
\showURL{%
\tempurl}


\bibitem[Shenoy(2023)]%
        {chatgotgraph}
\bibfield{author}{\bibinfo{person}{Varun Shenoy}.}
  \bibinfo{year}{2023}\natexlab{}.
\newblock \bibinfo{title}{GraphGPT}.
\newblock
  \bibinfo{howpublished}{\url{https://github.com/varunshenoy/graphgpt}}.
\newblock


\bibitem[Sorokina and Cantu-Paz(2016)]%
        {10.1145/2911451.2926725}
\bibfield{author}{\bibinfo{person}{Daria Sorokina} {and} \bibinfo{person}{Erick
  Cantu-Paz}.} \bibinfo{year}{2016}\natexlab{}.
\newblock \showarticletitle{Amazon Search: The Joy of Ranking Products}. In
  \bibinfo{booktitle}{\emph{Proceedings of the 39th International ACM SIGIR
  Conference on Research and Development in Information Retrieval}} (Pisa,
  Italy) \emph{(\bibinfo{series}{SIGIR '16})}. \bibinfo{publisher}{Association
  for Computing Machinery}, \bibinfo{address}{New York, NY, USA},
  \bibinfo{pages}{459–460}.
\newblock
\showISBNx{9781450340694}
\urldef\tempurl%
\url{https://doi.org/10.1145/2911451.2926725}
\showDOI{\tempurl}


\bibitem[Tagliabue(2021)]%
        {10.1145/3460231.3474604}
\bibfield{author}{\bibinfo{person}{Jacopo Tagliabue}.}
  \bibinfo{year}{2021}\natexlab{}.
\newblock \showarticletitle{You Do Not Need a Bigger Boat: Recommendations at
  Reasonable Scale in a (Mostly) Serverless and Open Stack}. In
  \bibinfo{booktitle}{\emph{Proceedings of the 15th ACM Conference on
  Recommender Systems}} (Amsterdam, Netherlands) \emph{(\bibinfo{series}{RecSys
  '21})}. \bibinfo{publisher}{Association for Computing Machinery},
  \bibinfo{address}{New York, NY, USA}, \bibinfo{pages}{598–600}.
\newblock
\showISBNx{9781450384582}
\urldef\tempurl%
\url{https://doi.org/10.1145/3460231.3474604}
\showDOI{\tempurl}


\bibitem[Tagliabue(2022a)]%
        {jacopo_medium}
\bibfield{author}{\bibinfo{person}{Jacopo Tagliabue}.}
  \bibinfo{year}{2022}\natexlab{a}.
\newblock \bibinfo{title}{{Applied Research at Reasonable Scale}}.
\newblock
  \bibinfo{howpublished}{\url{https://medium.com/the-techlife/applied-research-at-reasonable-scale-8a74d2beed89}}.
\newblock
\newblock
\shownote{[Online; accessed 19-Feb-2023]}.


\bibitem[Tagliabue(2022b)]%
        {towardDataScienceJT}
\bibfield{author}{\bibinfo{person}{Jacopo Tagliabue}.}
  \bibinfo{year}{2022}\natexlab{b}.
\newblock \bibinfo{title}{MLOps without Much Ops}.
\newblock
\newblock
\urldef\tempurl%
\url{https://towardsdatascience.com/mlops-without-much-ops-d17f502f76e8}
\showURL{%
\tempurl}


\bibitem[Tagliabue et~al\mbox{.}(2023)]%
        {tagliabue2023reasonable}
\bibfield{author}{\bibinfo{person}{Jacopo Tagliabue}, \bibinfo{person}{Hugo
  Bowne-Anderson}, \bibinfo{person}{Ville Tuulos}, \bibinfo{person}{Savin
  Goyal}, \bibinfo{person}{Romain Cledat}, {and} \bibinfo{person}{David Berg}.}
  \bibinfo{year}{2023}\natexlab{}.
\newblock \bibinfo{title}{Reasonable Scale Machine Learning with Open-Source
  Metaflow}.
\newblock
\newblock
\showeprint[arxiv]{2303.11761}~[cs.LG]


\bibitem[Tagliabue et~al\mbox{.}(2021)]%
        {CoveoSIGIR2021}
\bibfield{author}{\bibinfo{person}{Jacopo Tagliabue}, \bibinfo{person}{Ciro
  Greco}, \bibinfo{person}{Jean-Francis Roy}, \bibinfo{person}{Federico
  Bianchi}, \bibinfo{person}{Giovanni Cassani}, \bibinfo{person}{Bingqing Yu},
  {and} \bibinfo{person}{Patrick~John Chia}.} \bibinfo{year}{2021}\natexlab{}.
\newblock \showarticletitle{SIGIR 2021 E-Commerce Workshop Data Challenge}. In
  \bibinfo{booktitle}{\emph{SIGIR eCom 2021}}.
\newblock


\bibitem[Tagliabue et~al\mbox{.}(2020)]%
        {tagliabue-etal-2020-grow}
\bibfield{author}{\bibinfo{person}{Jacopo Tagliabue}, \bibinfo{person}{Bingqing
  Yu}, {and} \bibinfo{person}{Marie Beaulieu}.}
  \bibinfo{year}{2020}\natexlab{}.
\newblock \showarticletitle{How to Grow a (Product) Tree: Personalized Category
  Suggestions for e{C}ommerce Type-Ahead}. In
  \bibinfo{booktitle}{\emph{Proceedings of The 3rd Workshop on e-Commerce and
  NLP}}. \bibinfo{publisher}{Association for Computational Linguistics},
  \bibinfo{address}{Seattle, WA, USA}, \bibinfo{pages}{7--18}.
\newblock
\urldef\tempurl%
\url{https://doi.org/10.18653/v1/2020.ecnlp-1.2}
\showDOI{\tempurl}


\bibitem[Tay et~al\mbox{.}(2022)]%
        {tay2022transformer}
\bibfield{author}{\bibinfo{person}{Yi Tay}, \bibinfo{person}{Vinh~Q. Tran},
  \bibinfo{person}{Mostafa Dehghani}, \bibinfo{person}{Jianmo Ni},
  \bibinfo{person}{Dara Bahri}, \bibinfo{person}{Harsh Mehta},
  \bibinfo{person}{Zhen Qin}, \bibinfo{person}{Kai Hui}, \bibinfo{person}{Zhe
  Zhao}, \bibinfo{person}{Jai Gupta}, \bibinfo{person}{Tal Schuster},
  \bibinfo{person}{William~W. Cohen}, {and} \bibinfo{person}{Donald Metzler}.}
  \bibinfo{year}{2022}\natexlab{}.
\newblock \bibinfo{title}{Transformer Memory as a Differentiable Search Index}.
\newblock
\newblock
\showeprint[arxiv]{2202.06991}~[cs.CL]


\bibitem[{Techcrunch}(2021)]%
        {AlgoliaRound}
\bibfield{author}{\bibinfo{person}{{Techcrunch}}.}
  \bibinfo{year}{2021}\natexlab{}.
\newblock \bibinfo{booktitle}{\emph{Search API startup Algolia raises \$150
  million at \$2.25 billion valuation}}.
\newblock
\urldef\tempurl%
\url{https://techcrunch.com/2021/07/28/search-api-startup-algolia-raises-150-million-at-2-25-billion-valuation/}
\showURL{%
\tempurl}


\bibitem[Tsagkias et~al\mbox{.}(2020)]%
        {Tsagkias2020ChallengesAR}
\bibfield{author}{\bibinfo{person}{Manos Tsagkias},
  \bibinfo{person}{Tracy~Holloway King}, \bibinfo{person}{Surya Kallumadi},
  \bibinfo{person}{Vanessa Murdock}, {and} \bibinfo{person}{Maarten de Rijke}.}
  \bibinfo{year}{2020}\natexlab{}.
\newblock \showarticletitle{Challenges and Research Opportunities in eCommerce
  Search and Recommendations}. In \bibinfo{booktitle}{\emph{SIGIR Forum}},
  Vol.~\bibinfo{volume}{54}.
\newblock


\bibitem[Wang et~al\mbox{.}(2015)]%
        {wang-etal-2015-building}
\bibfield{author}{\bibinfo{person}{Yushi Wang}, \bibinfo{person}{Jonathan
  Berant}, {and} \bibinfo{person}{Percy Liang}.}
  \bibinfo{year}{2015}\natexlab{}.
\newblock \showarticletitle{Building a Semantic Parser Overnight}. In
  \bibinfo{booktitle}{\emph{Proceedings of the 53rd Annual Meeting of the
  Association for Computational Linguistics and the 7th International Joint
  Conference on Natural Language Processing (Volume 1: Long Papers)}}.
  \bibinfo{publisher}{Association for Computational Linguistics},
  \bibinfo{address}{Beijing, China}, \bibinfo{pages}{1332--1342}.
\newblock
\urldef\tempurl%
\url{https://doi.org/10.3115/v1/P15-1129}
\showDOI{\tempurl}


\bibitem[Wang et~al\mbox{.}(2021)]%
        {Wang2021}
\bibfield{author}{\bibinfo{person}{Yaxuan Wang}, \bibinfo{person}{Hanqing Lu},
  \bibinfo{person}{Yunwen Xu}, \bibinfo{person}{Rahul Goutam},
  \bibinfo{person}{Yiwei Song}, {and} \bibinfo{person}{Bing Yin}.}
  \bibinfo{year}{2021}\natexlab{}.
\newblock \showarticletitle{QUEEN: Neural query rewriting in e-commerce}. In
  \bibinfo{booktitle}{\emph{The Web Conference 2021}}.
\newblock
\urldef\tempurl%
\url{https://www.amazon.science/publications/queen-neural-query-rewriting-in-e-commerce}
\showURL{%
\tempurl}


\bibitem[Yan et~al\mbox{.}(2018)]%
        {Yan2018APD}
\bibfield{author}{\bibinfo{person}{Yan Yan}, \bibinfo{person}{Zitao Liu},
  \bibinfo{person}{Meng Zhao}, \bibinfo{person}{Wentao Guo},
  \bibinfo{person}{Weipeng~P. Yan}, {and} \bibinfo{person}{Yongjun Bao}.}
  \bibinfo{year}{2018}\natexlab{}.
\newblock \showarticletitle{A Practical Deep Online Ranking System in
  E-commerce Recommendation}. In \bibinfo{booktitle}{\emph{ECML/PKDD}}.
\newblock


\bibitem[Yu et~al\mbox{.}(2020)]%
        {10.1145/3366424.3386198}
\bibfield{author}{\bibinfo{person}{Bingqing Yu}, \bibinfo{person}{Jacopo
  Tagliabue}, \bibinfo{person}{Ciro Greco}, {and} \bibinfo{person}{Federico
  Bianchi}.} \bibinfo{year}{2020}\natexlab{}.
\newblock \showarticletitle{“An Image is Worth a Thousand Features”:
  Scalable Product Representations for In-Session Type-Ahead Personalization}.
  In \bibinfo{booktitle}{\emph{Companion Proceedings of the Web Conference
  2020}} (Taipei, Taiwan) \emph{(\bibinfo{series}{WWW '20})}.
  \bibinfo{publisher}{Association for Computing Machinery},
  \bibinfo{address}{New York, NY, USA}, \bibinfo{pages}{461–470}.
\newblock
\showISBNx{9781450370240}
\urldef\tempurl%
\url{https://doi.org/10.1145/3366424.3386198}
\showDOI{\tempurl}


\bibitem[Zhang and Chen(2018)]%
        {Zhang2018ExplainableRA}
\bibfield{author}{\bibinfo{person}{Yongfeng Zhang} {and} \bibinfo{person}{Xu
  Chen}.} \bibinfo{year}{2018}\natexlab{}.
\newblock \showarticletitle{Explainable Recommendation: A Survey and New
  Perspectives}.
\newblock \bibinfo{journal}{\emph{Found. Trends Inf. Retr.}}
  \bibinfo{volume}{14} (\bibinfo{year}{2018}), \bibinfo{pages}{1--101}.
\newblock


\bibitem[Zheng et~al\mbox{.}(2017)]%
        {zheng-etal-2017-joint}
\bibfield{author}{\bibinfo{person}{Suncong Zheng}, \bibinfo{person}{Feng Wang},
  \bibinfo{person}{Hongyun Bao}, \bibinfo{person}{Yuexing Hao},
  \bibinfo{person}{Peng Zhou}, {and} \bibinfo{person}{Bo Xu}.}
  \bibinfo{year}{2017}\natexlab{}.
\newblock \showarticletitle{Joint Extraction of Entities and Relations Based on
  a Novel Tagging Scheme}. In \bibinfo{booktitle}{\emph{Proceedings of the 55th
  Annual Meeting of the Association for Computational Linguistics (Volume 1:
  Long Papers)}}. \bibinfo{publisher}{Association for Computational
  Linguistics}, \bibinfo{address}{Vancouver, Canada},
  \bibinfo{pages}{1227--1236}.
\newblock
\urldef\tempurl%
\url{https://doi.org/10.18653/v1/P17-1113}
\showDOI{\tempurl}


\end{thebibliography}

\appendix

\section{Implementation notes}
\label{sec:appendix}

As the novelty of our proposal does not lie in new classifiers or NLP pipelines, we briefly expand here the implementation strategies sketched in Section \ref{sec:parsing}. We count as a strength of the approach that tried-and-tested and off-the-shelf techniques can be successfully used to start: any improvements to the below methods will make the parser even better. 

\subsection{Product representation}

Once basic tags (COLOR, BRAND, etc.) are defined as the building blocks of the knowledge base \textit{and} the logical forms, we need to know where and how each of these attributes can be found starting from the product catalogs. We favor a declarative approach, where tags are associated with strategies, that get executed in series when parsing the catalog. For example, Table~\ref{tab:parsingConfig} shows how three tags can be extracted from Shop A.

\begin{table}[h]
\centering
\begin{tabular}{p{0.2\linewidth}p{0.2\linewidth}p{0.42\linewidth}}
\hline
\textbf{Tag} & \textbf{Type} & \textbf{Strategy}\\
\hline
\textbf{Brand} & Config & \textit{Manufacturer} \\
\textbf{Color} & Model & CLIP-based classifier \\
\textbf{Product} & Heuristic & First noun in \textit{Description} overlapping with \textit{Category}\\
\hline
\end{tabular}
\caption{\label{tab:parsingConfig}
\textbf{Building a knowledge base for Shop A.} Three sample strategies for building symbolic product representations using naming conventions, machine learning, and domain specific heuristics.  
}
\end{table}

We first have \textit{configuration} strategy, which just points to the column in the catalog that contains the attribute (typical for brands, prices etc.); this leverages the structured nature of catalogs, which is a huge simplifying factor when considering product search \textit{vis-à-vis} web search. We then have a \textit{model} strategy, which relies on machine learning to accomplish tagging; finally we have a \textit{heuristic} strategy, building on domain knowledge and catalog specifics. 

When discussing scaling B2B product search across deployments, it's important to realize different strategies have different levels of granularity. Configurations are set \textit{per shop} and they are deterministic; models can typically be trained across shops (for entire industries for example) and can leverage the latest zero-shot classifiers in case no label is wanted / needed \cite{Naturearticle} \textit{heuristics} are more case specifics, but in our experience they have some degree of re-use: moreover, heuristics can be used to train new classifiers (using for example weak supervision \cite{Ratner2017SnorkelRT}), which will in turn reduce the use of heuristics.

Importantly, the very recent progress on large language models promises to greatly simplify the actual building of a structured knowledge representation, offering even zero-shot graph building from text \cite{chatgotgraph}. While LLMs are still too slow and somehow not understood enough to be directly involved in the runtime query path, they are definitely well suited to speed up the offline component of our method (Fig.~\ref{fig:teaser}, section 1 and 2 from the left).

\end{document}